\documentclass[aps,pra,twocolumn,groupedaddress,floatfix]{revtex4-1}
\usepackage{blindtext,amsmath}
\usepackage{graphicx,enumerate}
\usepackage{float}
\usepackage[export]{adjustbox}
\usepackage[caption=false]{subfig}

\newcommand{\gbb}{g_\mathrm{BB}}
\newcommand{\gib}{g_\mathrm{IB}}
\newcommand{\mred}{m_\mathrm{r}}

\newcommand{\ad}{\hat{a}^{\dagger}}
\newcommand{\oa}{\hat{a}}

\newcommand{\bk}{\mathbf{k}}
\newcommand{\bkd}{\mathbf{k'}}

\newcommand{\odphi}{\hat{\phi}^\dagger(\mathbf{r}) }
\newcommand{\ophi}{\hat{\phi}(\mathbf{r}) }

\usepackage{color}

\begin{document}
\title{The effect of boson-boson interaction on the Bipolaron formation}
\author{J. Jager}
\author{R. Barnett}
\affiliation{Department of Mathematics, Imperial College London, London SW7 2AZ, United Kingdom}
\begin{abstract}
Impurities immersed into a surrounding ultra-cold Bose gas experience interactions mediated by the surrounding many-body environment. If one focuses on two impurities that are sufficiently close to each other, they can form a bipolaron pair. Here, we discuss how the standard methods based on linearizing the condensate field lead to results only valid in the weak coupling regime and for sufficiently large impurity separations. We show how those shortcomings can be remedied within the Born-Oppenheimer approximation by accounting for boson-boson interactions already on the mean-field level.
\end{abstract}

\maketitle

%%%%%%%%%%%%%%%%%%%%%%%%%%%%%%%%%
\section{Introduction}
%%%%%%%%%%%%%%%%%%%%%%%%%%%%%%%%%
The interaction between  an impurity and a surrounding many-body environment can lead to the formation of a quasiparticle called a polaron \cite{Landau1933,Pekar1946}. 
When multiple impurities are present, exchange interactions, also mediated by
the surrounding environment, can lead to impurity-impurity bound states known as bipolarons.
Such exchange mediated interactions are ubiquitous in physical systems, being relevant for Cooper pairs in superconductors \cite{Cooper1956} and quark-gluon interactions \cite{Peskin:1995ev}. 
In the solid-state context, lattice phonon vibrations are responsible for the mediated interactions.
The resulting bipolarons may play 
 a role in high-$T_c$ superconductivity \cite{Mott1993,Alexandrov1992} and are also a vital ingredient for understanding the electric conductivity of polymers \cite{Bredas1985,Glenis1993}.  

In more recent years, neutral atoms immersed in ultra-cold quantum gases  have provided an excellent platform to investigate the physics of polarons. 
Being highly tuneable via Feshbach resonances \cite{Chin2010a}, such systems allow access to novel regimes.  
Here the density fluctuations of the ultracold gas mediate the interaction between impurities which can result in a bound state as illustrated in FIG.~\ref{fig:Binding_energy}.
Using ultracold quantum gases,
the Fermi-polaron  has been investigated in a several experiments  \cite{Schirotzek2009,Zhang2012,Kohstall2012,Koschorreck2012,Scazza2017,Cetina2015,Cetina2016,Ness2020,Yan2019a} and in recent years the experimental progress in  Bose-polarons has also made considerable advances \cite{Catani2012,Yan2019,Jorgensen,Skou2021, Hu}.
A common starting point for describing impurities in an ultracold Bose gas is the linearized Fr\"ohlich model \cite{Grusdt2015a,Grusdt2015a,Rath2013,Li2014}. Despite its applicability to the weak coupling regime, it is known from the single impurity case \cite{Ardila2015,Shchadilova2016}, that the Fr\"ohlich model becomes inadequate when applied to strongly interacting impurities. {\color{black} A natural next step is to consider the extended Fr\"ohlich model which systematically accounts for impurity-boson interactions of higher order, i.e. by retaining second order phonon impurity process while still neglecting phonon-phonon interaction \cite{Shchadilova2016}.}
Although the extended Fr\"ohlich model, has been applied with considerable success 
to dynamical phenomena and describing repulsive and weakly attractive interactions
\cite{Shchadilova2016,Drescher2019,Ashida2018,Dzsotjan2020,Lausch2018,Ichmoukhamedov2019},
it too possesses some significant shortcomings. For instance an instability can form due to the emergence of a bound state. The extended Fr\"ohlich model predicts that an infinite number of bosons populates this energetically low-lying bound state which is typically unphysical \cite{Shchadilova2016}. That is, in a realistic interacting Bose gas, the high occupancy of the bound state is balanced by the boson-boson repulsion \cite{Schmidt,PhysRevLett.127.033401}. 
%In \cite{Camacho-Guardian2018} it was demonstrated that the Yukawa potential is not entirely accurate and is only valid for weak couplings and sufficiently large separation between the two impurities. Further insight for the 1D case was gained in \cite{Will2021}, where the boson-boson interaction was already accounted for at the mean-field level, and excellent agreement with quantum Monte Carlo results was reported. {\color{black} The importance of accounting for the condensate deformation at the mean-field level to describe a single impurity in 1D has previously been emphasized
%\cite{Jager2020,Volosniev2017,Mistakidis2019,Panochko2019}.}
%Building on the results of the lower dimensional counterpart, recent efforts have shown that also in three dimensions, the polaron formation can be described accurately by taking the boson-boson interaction into account directly \cite{Schmidt,Drescher2020,Guenther2020}.
Describing the interaction between two neutral impurities immersed in a Bose gas is crucial for understanding the interplay between several impurities. Here, the Fr\"ohlich model predicts, within the Born-Oppenheimer approximation,  an attractive Yukawa potential between two impurities in 3D \cite{Naidon2018}.
In \cite{Camacho-Guardian2018} it was noted however that the Yukawa potential is not entirely accurate, being only valid for weak couplings and sufficiently large impurity separation. Building on the single impurity case, one therefore might expect that the results obtained from the Fr\"olich model for weak couplings can be improved upon in a straightforward way by including higher-order phonon impurity scattering terms.
However, we will show that if one proceeds in a naive manner for two impurities, this can lead to unphysical divergences in the ground state energy due to the bound state formation between the two impurities and the excitations of the Bose gas, something that has also been demonstrated in \cite{Panochko2022}. In contrast to the single impurity case, this occurs for attractive and repulsive impurity-boson scattering lengths. The mechanisms leading to this bound state are similar to those leading to the bound state formed between two localized potentials known from standard quantum mechanics \cite{Albeverio2004}.

In this work, we present a conceptually simple and physically intuitive  model to address the bipolaron problem. This model constitutes a good starting point for more advanced treatments and also rectifies the shortcomings of the (extended) Fr\"ohlich model {\color{black} when considering the bipolaron problem}. We start by introducing the full microscopic Hamiltonian. 
We proceed by linearizing the model and integrating out the phononic degrees of freedom which leads to the Yukawa potential. We then discuss why the Yukawa potential is inadequate and also outline why some of the standard methods used to go beyond the Fr\"ohlich model in the single impurity case do not generalize in a straightforward manner.
We then show how those problems can be remedied in a conceptually simple and intuitive way by accounting for boson-boson interaction at the mean-field level, in line with previous treatments of bipolarons in 1D \cite{Will2021,Dehkharghani2018} and single polarons \cite{Schmidt,Guenther2020,Drescher2020,Jager2020,Mistakidis2019a,Brauneis2021,Koutentakis2021,Mistakidis2019,PhysRevLett.127.033401}. 
This is done by applying the Lee-Low-Pines transformation \cite{Lee1953} and transforming to the center of mass coordinates for the two impurities. This brings the Hamiltonian into a form amenable to the Born-Oppenheimer (BO) approximation. We proceed by minimizing the resulting Gross-Pitaevskii (GP) energy functional. This leaves us with an effective Schr\"odinger equation for the two impurities with which we determine conditions for a bound state to occur.

%%%%%%%%%%%%%%%%%%%%%%%
\section{The Model}
%%%%%%%%%%%%%%%%%%%%%%%
Our starting point is a microscopic theory describing
two impurities coupled to a surrounding Bose gas, consisting of $N$ particles in a box of volume $V$ with periodic boundary conditions. Such a system is described by the Hamiltonian 
\begin{align} 
 \label{eq:Hamiltonian}
     \hat{\mathcal{H}} &= \int \mathrm{d}^3r\, \odphi \bigg(-\frac{ \nabla^2}{2m} +\frac{\gbb}{2}\odphi\ophi-\mu\nonumber \\
     &+V(\mathbf{r}-\hat{\mathbf{R}}_1)+V(\mathbf{r}-\hat{\mathbf{R}}_2)\bigg)\ophi +\frac{\hat{\mathbf{P}}_1^2+\hat{\mathbf{P}}_2^2}{2M}.
\end{align}
Here we set $ \hbar = 1$ and $m$ ($M$) denotes the mass of the bosons
(impurity atoms), $\hat{\phi}(x)$ is the bosonic
field operator describing the Bose gas, $\gbb$ ($\gib$) is the boson-boson (boson-impurity) interaction strength, 
{\color{black}$\hat{\mathbf{X}}_{1,2}$ ($\hat{\mathbf{P}}_{1,2}$)} denotes the
position (momentum) operator of the impurities, and 
$\mu$ is the chemical potential of the Bose gas.
 The interaction between the impurities and the condensate is modelled by the interaction potential $V(\mathbf{r})$; most linearized treatments rely on employing a contact potential
  {\color{black} $V_\delta(\mathbf{r})= \gib \delta({\bf r})$ }
  \cite{Naidon2018,Camacho-Guardian2018}.
 As is known for such models, when keeping the full Hamiltonian and applying a contact interaction for the impurity-boson interaction and the boson-boson interaction simultaneously, the Hamiltonian only admits zero energy (bi)polaron solutions \cite{Guenther2020}. Thus when working with the non-linearized model in the Born-Oppenheimer approximation {\color{black} at least one of the two interactions has to be chosen to be of finite range. In this work, we employ a finite-range potential  for the impurity-boson interaction. For the boson-boson interaction we still employ a contact interaction. We choose the widely-used Gaussian pseudo-potential
 \begin{align}
      V_{\rm G}(r) = -V_0e^{- \frac{r^2}{L^2}},
 \end{align}
 with depth $V_0$ and range $L$ and also compare the results to the soft van-der-Waals potential
 \begin{align}
      V_{\rm vdw}(r) = -V_0 \frac{L^6}{r^6+L^6} \, .
 \end{align}}
 {\color{black} The connection to the s-wave scattering length $a_{\rm IB}$ and the effective range $r_{\rm eff}$ can be made by numerically solving the two-body Schr\"odinger equation \cite{Jeszenszki2018,Stoof2009}.
For a spherical potential $u(r)$ satisfies the (radial) differential equation $ \left( -\frac{\mathrm{d}^2}{\mathrm{d}r^2} +2m_\mathrm{r}V(r) +k^2 \right)u_k(r) = 0,$
where $m_r = mM/(m+M)$ and the boundary conditions are $u(0) = 0$ and $u'(0)=1$. 
By solving for $u_k(r)$ one can now extract the phase shift $\delta_0(k)$, which ultimately determines the scattering length and effective range via
\begin{align}
    k\cot{\delta_0(k)} = -\frac{1}{a_{\rm IB}} +\frac12 r_{\rm eff} k^2 +\mathcal{O}(k^4) \, .
\end{align}
This relation can be used to make the connection to the contact potential used in the linearized case.}

To conclude this section, we introduce the relative coordinates and apply a unitary transformation to eliminate the center of mass degrees of freedom. Starting with Eq.~(\ref{eq:Hamiltonian}), we transform into the center of mass frame and denote the relative position (momentum) of the impurities by $\hat{\mathbf{R}}$ ($\hat{\mathbf{P}}$) and the center of mass position (momentum) by $\hat{\mathbf{r}}_{\rm I}$ ($\hat{\mathbf{p}}_{\rm I}$). Subsequently we apply a Lee-Low-Pines transformation $\hat{U} = \exp(i\hat{\mathbf{r}}_{\rm I}\cdot\hat{\mathbf{P}}_\mathrm{B})$ ,where $\hat{\mathbf{P}}_\mathrm{B} = -i\int \mathrm{d}^dr\, \odphi \nabla \ophi $ is the total momentum of the Bose gas. This eliminates the center of mass coordinate \cite{Will2021,Lee1953} and we arrive at the following Hamiltonian
\begin{align} 
 \label{eq:HamiltonianLLP}
     &\hat{\mathcal{H}}= \frac{:\left(\mathbf{p}-\hat{\mathbf{P}}_\mathrm{B}\right)^2:}{4M}+\frac{\hat{P}^2}{M} +\int  \mathrm{d}^3r\, \odphi \bigg(\frac{-\nabla^2}{2m_r}\\\nonumber&+\frac{\gbb}{2}\odphi\ophi -\mu +V(\mathbf{r}+\hat{\mathbf{R}}/2)+ V(\mathbf{r}-\hat{\mathbf{R}}/2)\bigg)\ophi.
\end{align}
Here,
 $m_r$ is the reduced mass and $\mathbf{p}$ is the total momentum, which is a conserved quantity and therefore can be replaced by a {\color{black} real} number. 
{Throughout our calculations we set $p=0$ since we focus on systems at rest to obtain the mediated interaction.}
 One might notice that we are neglecting direct impurity-impurity interactions in our considerations. This is strictly speaking only allowed when the impurities are well separated. As will be further explained, the range of the bare impurity-impurity interaction will usually be much smaller than the range of the mediated potential.
The standard procedure is to linearize the field operators and subsequently perform a Bogoliubov rotation, resulting in the (extended) Fr\"ohlich model. The following section will briefly outline how to retrieve these results by linearizing only the density and neglecting phase-density interactions.

%%%%%%%%%%%%%%%%%%%

\section{{Linearized theory}}

%%%%%%%%%%%%%%%%%%%

In this section, we address the bipolaron problem utilizing a path-integral approach, which is expanded in density fluctuations. Though the resulting expressions can be obtained directly from the Fr\"ohlich model, the path integral approach gives a clearer picture of how the interaction is mediated by the density fluctuations of the condensate. Furthermore, it demonstrates that the neglected boson-boson interaction is the root cause of the shortcomings in predicting the mediated interactions. 
We start by rewriting the field operators as $\ophi = \sqrt{n_0 +\delta\hat{\rho}(\mathbf{r)}}e^{i\hat{\theta}(\mathbf{r})}$, where $n_0=\mu/\gbb$. After performing this redefinition, dropping terms of order higher than quadratic in $\delta \rho$ and $\partial_i \theta$, we arrive at the imaginary-time action 
\begin{align}
    S = \int \mathrm{d}\tau\,\Big\{\int \mathrm{d}^3r\Big[ \delta\rho \partial_\tau \theta + \frac{ n_0(\nabla \theta)^2}{2m_{\rm r}} +\frac{1}{2}\delta\rho\Big(\frac{-\nabla^2}{4m_{\rm r}n_0}\\ \nonumber+\gbb\Big)\delta\rho+\gib\left(\delta\rho(\mathbf{R}/2)+\delta\rho(-\mathbf{R}/2)\right)\Big] +\frac{ \mathbf{P}^2}{M}\Big\}
\end{align}
It is now straightforward to first integrate out the density and subsequently the phase, which leaves us with an effective action for the impurities (see \cite{Ichmoukhamedov2019,Tempere} for similar calculations for the Bose polaron)
\begin{align}
    S = \sum_n\,\Big\{\frac{\mathbf{P}^2}{M} -\frac{\gib^2}{(2\pi)^3}\int \mathrm{d}^3k\,\frac{n_0 e_k\cos{(\mathbf{k}\cdot \mathbf{R}/2}) }{\Omega_k^2+\omega^2_n} \Big\}
\end{align}
%\begin{align}
%S =& \int \mathrm{d}^4{\tilde{p}} \, \psi_{\tilde{p}}^* (-i \omega_n +\frac{p^2}{2M})\psi_{\tilde{p}} \\ \nonumber &+\frac{1}{2} \int\mathrm{d}^4\tilde{p}\mathrm{d}^4\tilde{p}'\mathrm{d}^4\tilde{q} \, \psi_{\tilde{p}+\tilde{q}}^*\psi_{\tilde{p}'-\tilde{q}}^* V_{\rm eff}(\tilde{q}) \psi_{\tilde{p'}}\psi_{\tilde{p}}.
%\end{align}
{where  $\omega_n$ are Matsubara frequencies,  $e_q = q^2/2m_r$ is the energy of the free boson and $\Omega_q = \sqrt{\frac{e_q}{2}(e_q+2n_0\gbb)}$ is the Bogoliubov dispersion.}  This leads to the mediated interaction
\begin{align}
     V_{\rm eff}(\omega_n,\mathbf{R}) = -\frac{\gib^2}{(2\pi)^3}\int \mathrm{d}^3k\,\frac{n_0 e_k\cos{(\mathbf{k}\cdot \mathbf{R}/2}) }{\Omega_k^2+\omega^2_n} 
\end{align}
where by evaluating the momentum integral and {applying the Born-Oppenheimer approximation, which allows us to take $\omega_n =0$,}
one can obtain the mediated interaction in real space. The calculations are straightforward and yield the Yukawa potential
%\begin{align}
%%    V_{BP}(R) = \begin{cases}
%-\frac{4 \pi a^2_\mathrm{IB}n_0}{m_r R}e^{-2\sqrt{2}R/\tilde{\xi}} &d=3\\
%-4 \pi^2/(m_r \ln(1/n a_{\rm IB}^2)^2) K_0(R/\tilde{\xi}) \quad  &d=2\\
%\end{cases},
%\end{align}
\begin{align} \label{eq:Yukawa}
    {V}^{\rm \delta}_{\mathrm{BP}}(R)=-\frac{4 n_0\pi a^2_\mathrm{IB}}{m_rR}e^{-\sqrt{2} R/\xi}
\end{align}
where we have used the usual relation $g_\mathrm{IB} = \frac{2\pi}{m_{\rm r}} a_\mathrm{IB}$ 
and introduced the healing length $\xi = \frac{1}{\sqrt{2\gbb n_0m_{\rm r}}}$.
We note that for heavy impurities and moderate couplings, one can find the ground state energy of the biplaron by solving the resulting Schr\"odinger equation. For heavy impurities, one can use the generalized parametric Nikiforov–Uvarov method to calculate approximate eigenenergies for the Yukawa potential \cite{Potential2012}, which results in the ground state energy
\begin{equation} \label{eqn:Yuka_eng} 
    E^{\rm \delta}_{BP} = -4 \pi^2 \frac{Mn_0^2a_\mathrm{IB}^4}{m_{\rm r}^2}.
\end{equation}
We note that this bound state only exists when $ |a_\mathrm{IB}| \geq \sqrt{\frac{m_{\rm  r}}{\sqrt{2}Mn_0\pi\xi}}$ \cite{Potential2012}. One can already see a major shortcoming of this approach, namely the bound state energy scales linearly in $M$ and the bipolaron energy diverges for $M\rightarrow\infty$. This is unphysical. The diverging energy can be traced back to the fact that the Yukawa potential is unbounded from below, and for $M \rightarrow\infty$, the kinetic energy becomes irrelevant. {\color{black} The unboundedness of the mediated interaction is a direct consequence of the delta function constituting a zero-range potential and therefore requires regularization. The regularization employed for the delta function regularizes the scattering for each impurity separately and is valid as long as the particles have non-zero separation, but breaks down when the impurities sit on top of each other, which effectively constitutes a single impurity with twice the bare interaction.}
Hence the ground-state energy becomes proportional to the minimum of the potential, which is $-\infty$ for the Yukawa potential. Additionally this treatment predicts a divergence at the Feshbach resonance.
{ \color{black}
Repulsive boson-boson interaction prevents an infinite number of bosons from attaching to the impurity, due to internal pressure arising from an increased number of bosons in a finite volume and we thus we do not expect such a divergence when accounting for boson-boson interactions. }
 \begin{figure}
	\centering 
	\includegraphics[scale = .99]{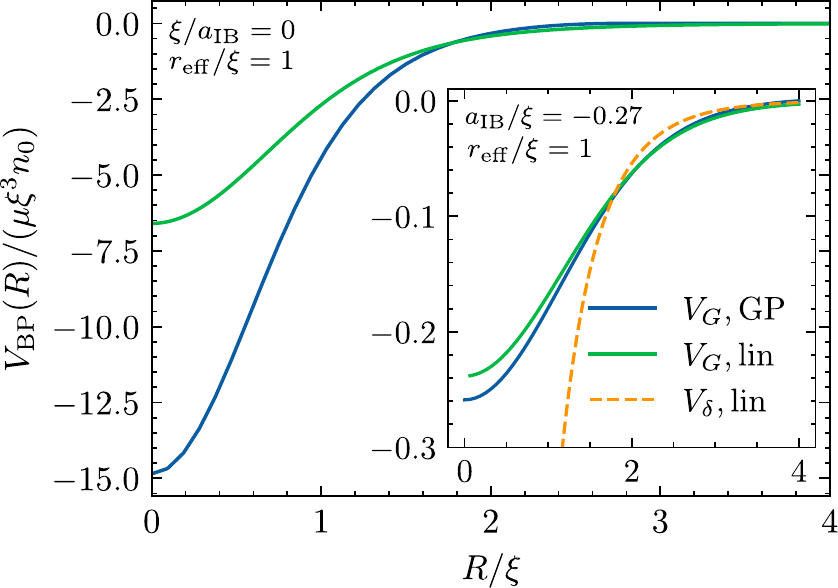}
	\caption{\color{black} The mediated potential for the two impurities near the Feshbach resonance. We can see that the transition for a more realistic pseudo-potential is smooth across the resonance. In contrast the diverging scattering length leads to a breakdown of the Yukawa predictions. It also becomes clear that for larger $a_\mathrm{IB}$, the Fr\"ohlich model, even with a UV regulated pseudo-potential, becomes inadequate.
	The inset shows the mediated interaction potential for small  $a_\mathrm{IB}$ (corresponding to small $V_0$). Here we can see that the linearized theory and the GP results agree fairly well. Additionally it should be noted that the Yukawa potential obtained by using a regularized contact interaction is only accurate for large separations.  }
	\label{fig:Int_pot} 
\end{figure}
In principle, one can improve upon these results by expanding the action perturbatively and resumming certain classes of diagrams. However, as shown in \cite{Schmidt} for the case of a single impurity, this is strictly speaking beyond the validity of the model and can lead to unphysical results near the scattering resonance due to the breakdown of the model associated with the bound state formation. In Appendix A, we show with the help of the extended Fr\"ohlich model for two impurities, that this can be problematic and can lead to divergences in the mediated potential. The idea is simple, in analogy with the case of a single particle interacting with two delta potentials (see \cite{Albeverio2004} and Appendix B), a bound state can form between the excitations and the two impurities. This bound state is energetically favorable and, without phonon-phonon interaction preventing an accumulation in this state, the condensate breaks down. To alleviate those problems, one has to incorporate phonon-phonon interaction and employ an interaction potential with finite range. The following section describes how this can be done by considering the boson-boson interaction at the mean-field level.

 %\begin{figure}
%	\centering
%	\includegraphics[scale = 0.83]{a_combined.pdf}
%	\caption{a) The dependence of the scattering length $a_\mathrm{IB}$ on the microscopic amplitude of the potential $V_0$ (which is taken to be attractive throughout this work) for $L = \xi$. The two regions considered in the rest of the paper are indicated. In region $\mathrm{I}$ ($\mathrm{II}$) a negative (positive) scattering length is obtained. b) The dependence of the scattering length of the mediated potential on the microscopic amplitude of $V_0$, }
%	\label{fig:scattering} 
%\end{figure}

Repeating the above analysis using the Gaussian potential (instead of a contact potential), one finds
\begin{multline} \label{eq:Gauss}
        V^{\rm G,\, lin}_\mathrm{BP}(R)= -\frac{2V^2_0 L^6 \pi m }{R} \times\\ \int^\infty_0  \frac{\sin(qR) n_0 q}{q^2+2/\xi^2}\exp\left(-q^2L^2/2\right)\, \mathrm{d}q.
\end{multline}
Note that with this potential,  $V^\mathrm{Gauss}_\mathrm{BP}$ stays finite for small $R$. This can be understood by noting that the exponential cut-off {\color{black} $\exp\left(-q^2L^2/2\right)$} is an effective UV-regulator, which is absent in the case of a delta function potential. However, if one uses this scattering potential for the extended Fr\"ohlich model, the bound state problem will persist. Additionally, the model loses the appeal of being analytically tractable when including higher-order phonon terms.
{\color{black} To summarize, while both \eqref{eq:Yukawa} and\eqref{eq:Gauss} are obtained by linearizing the model only \eqref{eq:Yukawa} assumes a contact potential and is thus ill-defined for $R=0$} 

\section{Main methodology and results}
 \begin{figure}[t]
	\includegraphics[scale =0.95]{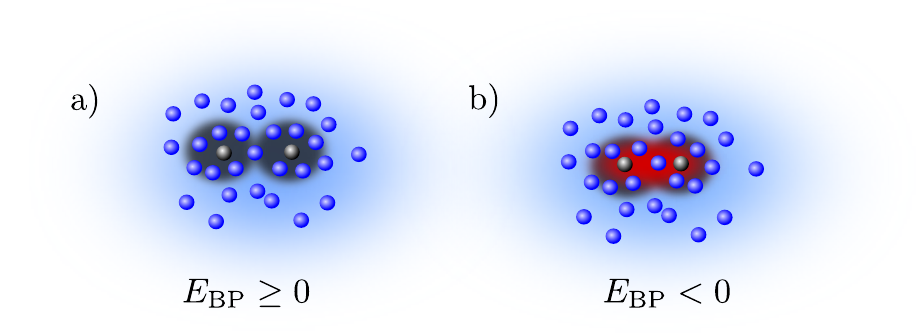}
	\includegraphics[scale =0.84]{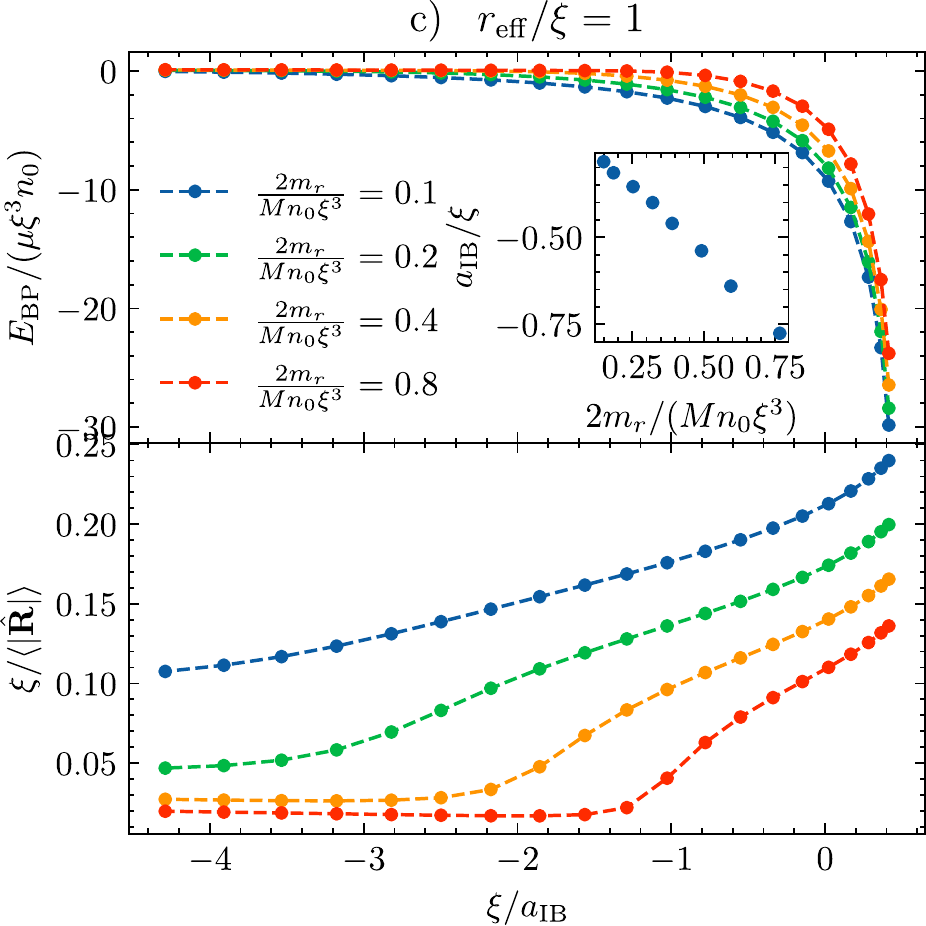}
	\caption{ \color{black} a) Schematic of bipolaron formation. If the interaction strength is weak or the impurities are very light, no bound state is formed, and only two polarons coexist. b) For strong enough interactions, a bound state called the bipolaron emerges.
	c) The binding energy and inverse separation of the bipolaron for {\color{black} different impurity kinetic energy scales }as a function of the inverse scattering length. The separation of the impurities is defined through $\langle|\mathbf{\hat{R}}| \rangle = \langle|\mathbf{\hat{R}}_1-\mathbf{\hat{R}}_2| \rangle $.The results are for $r_\mathrm{eff} = 1$ and obtained using $V_{\rm G}$. From the inverse separation it can clearly be seen when a bound state (the bipolaron) is formed. This threshold decreases with $\frac{m_{\rm r}}{M n_0\xi^3}$ and in fact becomes $0$ for $\frac{m_{\rm r}}{M n_0\xi^3} = 0$. In the inset we show the scattering length threshold after which a bound state is formed as a function of the mass ratio. \label{fig:Binding_energy} }
\end{figure}
In this section, we describe an approach which eliminates the difficulties encountered in
the previous section.
The Hamiltonian (\ref{eq:HamiltonianLLP}) will serve as the starting point for the mean-field treatment. The GP energy functional in the BO approximation that needs to be minimized to find the ground state can now be simply read off from (\ref{eq:HamiltonianLLP})
\begin{align}
    E(\phi,\mathbf{R}) =& \int \mathrm{d}^3r\, \Bigg\{ \frac{|\nabla\phi|^2}{2m_{\rm r}} +\frac{\gbb}{2}(|\phi|^2-n_0)^2 \\ \nonumber &+\left( {V}(\mathbf{r}+\mathbf{R/2})+{V}(\mathbf{r}-\mathbf{R/2})\right)|\phi|^2 \Bigg\}.
\end{align}
{\color{black}
To minimize the energy functional we have used the split-step Fourier algorithm in imaginary time}
\footnote{All calculations were performed in Cartesian coordinates, and to speed up the calculations, they were performed on GPUs using \texttt{CUDA.jl} \cite{besard2018juliagpu}.}.
This, in turn, allows us to calculate the mediated interaction through 
\begin{align}
    V^{\rm GP}_{\rm BP}(R) = E(R)-E_0-E(\infty),
\end{align}
where $E_0$ is the energy of the Bose gas without impurities and $E(\infty)$ is the energy of the two polarons at infinite separation and is subtracted to obtain the purely attractive part attributed to the bipolaron. {\color{black} Before discussing the main results we want to show that the problem can in fact be characterised by a few re-scaled parameters, which can then be used to interpret the results in terms of experimentally observable quantities. First we note that the chemical potential can be written as $\mu = \frac{4\pi a_\mathrm{BB}}{\mred} n_0$. By rescaling $\phi \rightarrow \phi \sqrt{n_0} $, $r\rightarrow r\xi$ and $V \rightarrow V/\mu$ we then find 
\begin{align}
   \frac{E(\phi,\mathbf{R})}{\mu\xi^3n_0} =& \int \mathrm{d}^3r\, \Bigg\{ |\nabla\phi|^2 +\frac{1}{2}(|\phi|^2-1)^2 \\ \nonumber &+\left( {V}(\mathbf{r}+\mathbf{R/2})+{V}(\mathbf{r}-\mathbf{R/2})\right)|\phi|^2 \Bigg\} \,.
\end{align}
%%%%%%%%%%%%%
Which shows that within the validity of of the c-field treatment our results are characterised only by the re-scaled energy, interaction strength, and impurity mass.}
%In FIG.~\ref{fig:scattering} a), we show the scattering length for different microscopic interaction amplitudes $V_0$, while keeping the potential range $L=\xi$ fixed throughout all of our calculations.
%We focus on (microscopically attractive) $V_0>0$ 
%on either side of the first resonance.
%{ Thus we can identify two relevant regimes, which are depicted in FIG.~\ref{fig:scattering} a). In area I (II), we find a negative (positive) scattering length for the impurity-boson interaction. For comparison  in FIG.~\ref{fig:scattering} b), we show the s-wave scattering length for the mediated impurity-impurity interaction.}

{In FIG.~\ref{fig:Int_pot} we show the shape of the mediated potential between two impurities for different $a_{\rm IB}$. The inset shows the comparison with the linearized model for weak coupling; here, we chose the s-wave scattering length of the Yukawa potential to match the scattering length of the Gaussian potential. For weak coupling, there is good quantitative agreement between the linearized model using a Gaussian pseudo-potential and the result obtained using the GP functional (see inset). We also observe that the Yukawa potential, which is obtained by employing a zero-range interaction, matches the behavior of the interaction potential with finite range for larger separations, indicating that the exact effective range of the potential is not highly relevant for the range of the mediated potential.
One main difference between a zero range interaction and a more realistic Gaussian interaction is that the mediated potential stays finite for $R=0$. A similar discrepancy between the Yukawa potential and the mediated potential was reported in \cite{Camacho-Guardian2018} using a scattering matrix approach.

In FIG.~\ref{fig:Int_pot} we also show the mediated interaction close to the Feshbach resonance. Here another shortcoming of the zero range scattering potential is revealed, namely close to the scattering resonance $a_\mathrm{IB}$ diverges, leading to infinite attraction, which is unphysical. The results obtained from the GP energy functional and the result obtained employing a Gaussian potential give a more realistic picture. Here, the mediated interaction changes less drastically across the Feshbach resonance. We also note that for larger $a_{\rm IB}$ (corresponding to larger $V_0$), the linearization approach becomes inadequate and significant deviation from the GP result can be observed. 
While the Fr\"ohlich model with Gaussian potential underestimates the interaction here, we note that it is not a priori clear whether the Fr\"ohlich model overestimates or underestimates the mediated potential. The two competing effects that the Fr\"ohlich model does not account for are (i) two and higher-order phonon impurity scattering processes, which lead to enhanced mediated impurity-impurity interaction and (ii) the boson-boson interaction, which damps the phonon exchange. {\color{black} We can see that changing $a_{\rm BB}$, while keeping all other parameters constant effectively results in re-scaling the impurity boson scattering length. Thus we move from the situation depicted in the inset of FIG.~\ref{fig:Int_pot} to the one shown in the main part of FIG.~\ref{fig:Int_pot}. This is exactly what one would expect, by noting that for large boson-boson interaction higher order phonon terms are damped out quickly and by neglecting the damping in the Fr\"ohlich model we overestimate the mediated interaction.}}
%In the two dimensional one has to be more careful at singularity close to zero by introducing the cut-off $\epsilon$ and one obtains
%\begin{align}
%    a^{2D}_{IB} = \lim_{r \rightarrow \infty, \epsilon \rightarrow 0} \left( 2r %\exp\left(- f(r,\epsilon)/(r f'(r,\epsilon)) -\gamma \right)\right),
%\end{align}
%where $f(r)$ satisfies
%\begin{align}
%\left(-\frac{\mathrm{d}^2}{\mathrm{d}r^2} %-\frac{1}{r}\frac{\mathrm{d}}{\mathrm{d}r}+\tilde{V} \right)f(r) =0.
%\end{align}
%The boundary conditions are $f(\epsilon)=1$ and $f'(\epsilon)=0$. Here, we not that the %scattering length for the two-dimensional case is always positive. 
  \begin{figure}
	\centering
		\includegraphics[scale = 1]{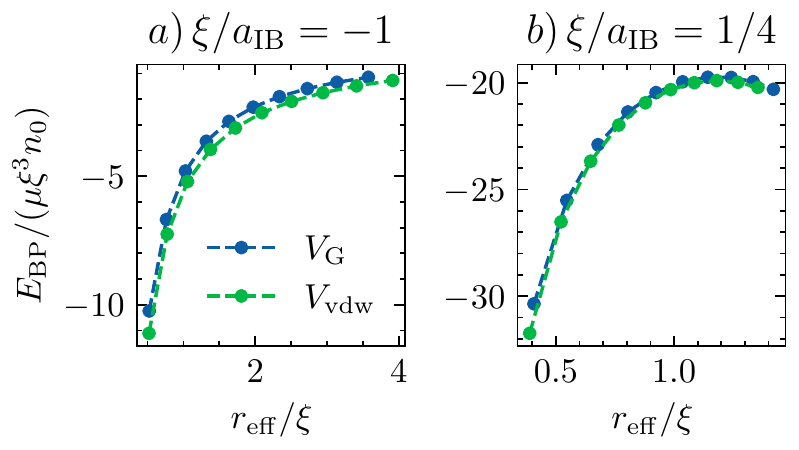}
		\includegraphics[scale =1]{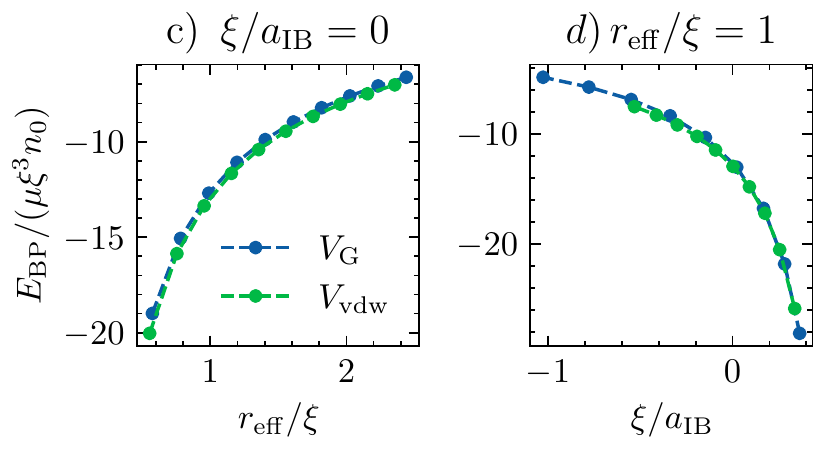}
	\caption{{\color{black} a),b),c): The bipolaron energy 
	 for different $a_\mathrm{IB}$ as a function of the effective range $r_{\rm eff}$.  In d), the  effective range is fixed and the bipolaron energy is plotted as a function of inverse scattering length.
	The plots demonstrate that the bipolaron energy is almost completely determined by the scattering length and the effective range and using different underlying potentials leads to similar results. All curves are for the $M= \infty$ case.} }
	\label{fig:comparison} 
\end{figure}

The bipolaron energy can be calculated by finding the ground-state of the resulting stationary Schr\"odinger equation.  We note that within the mean-field approximation for $p=0$, the wave function $\phi$ can always be chosen to be real. Therefore we do not have to consider the vector-potential typically arising within the Born-Oppenheimer approximation \cite{Wilczek1989}. Moreover, we observe that the equation is radially symmetric and that the ground-state will have zero angular momentum. Hence, we have to solve the following radial Schr\"odinger equation to obtain the bipolaron energy
{\color{black}
\begin{align}\label{eq:Schr}
    \left(-\frac{1}{M}\frac{\mathrm{d}^2}{\mathrm{dR^2}} + V^\mathrm{GP}_\mathrm{BP}(R)\right)u(R) = E_\mathrm{BP}u(R),
\end{align}}
with the boundary condition $u(0)=0$.
The results obtained are shown in FIG.~\ref{fig:Binding_energy}. 
{\color{black} We note that strictly speaking, our approach is only valid for $m/M \ll 1$. This corresponds to a vanishing impurity kinetic energy scale $ \frac{m_{\rm r}}{M n_0\xi^3}$, which serves as a control parameter for the Born-Oppenheimer approximation. }
It is notable that the dependence on the mass ratio is weak compared to the linearized case {(compare with (\ref{eqn:Yuka_eng}), where the energy scales linearly in the impurity mass)}, which can be explained by realizing that the effective potential stays finite. This can be understood by comparing kinetic energy to the potential energy. If the mass ratio becomes small, the kinetic energy becomes less important, and the solution of (\ref{eq:Schr}) will be localized around the minimum of the potential.
{\color{black}
Furthermore, we observe a critical $a_\mathrm{IB}$ after which a bipolaron characterized by $E_\mathrm{BP}<0$) is formed, see d, see also the inset of FIG.~\ref{fig:Binding_energy}.} In FIG.~\ref{fig:Binding_energy} {\color{black} this can also be clearly identified from the inverse separation of the impurities. We also note that the transition across the scattering resonance is smooth and the bipolaron binding energy further increases after crossing the scattering resonance. {\color{black} This can be understood by noting that the amplitude of the mediated potential $V^\mathrm{GP}_\mathrm{BP}$ increases further after crossing the resonance. Additionally, for heavy impurities the binding energy is approximately related to polaron energy through the approximate relationship $E_{BP} \approx E_{pol}(2V_0) -2E_{pol}(V_0)$, which becomes exact in the limit $M \rightarrow \infty$, since here the impurity kinetic energy becomes negligible and thus $\langle|\hat{\mathbf{R}}|\rangle \rightarrow 0$. Since in the regime after crossing the resonance the polaron energy scales faster than linearly \cite{Schmidt,PhysRevLett.127.033401} we expect the bipolaron binding energy to increase further across the resonance. We remark that the above argument relies on the validity of the GP treatment. In fact for $a_\mathrm{IB}>0$ two body bound states can appear that invalidate the GP treatment and while a detailed study of this regime is beyond the scope of this work it could be an interesting direction for further studies.}}

\begin{figure}
	\centering
		\includegraphics[scale = 0.9]{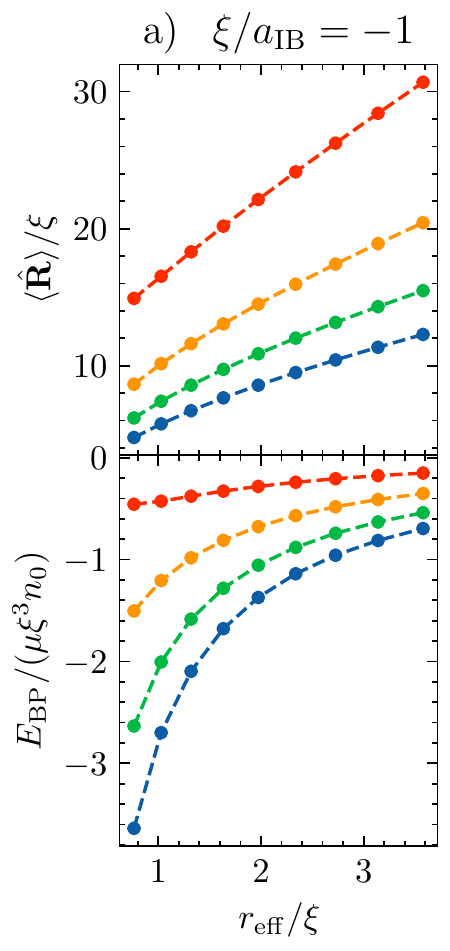}
		\includegraphics[scale =0.9]{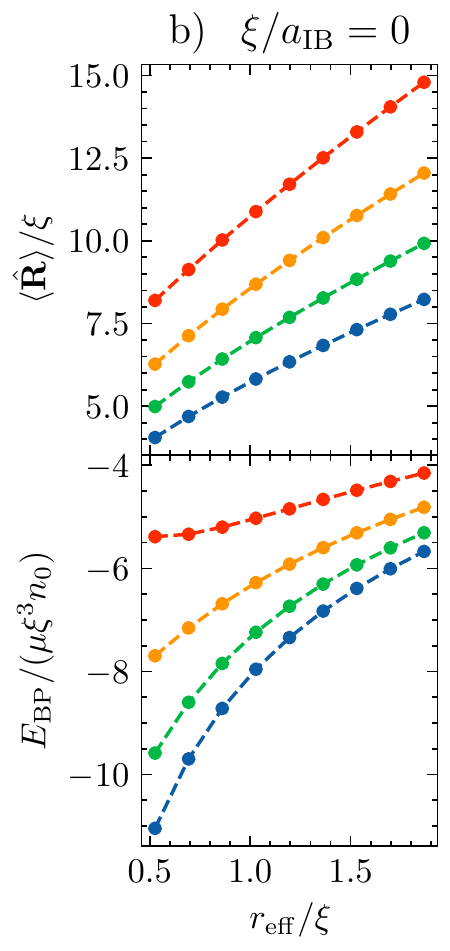}
	\caption{{\color{black}The effective range dependence of the bipolaron energy and size the for different mass ratios and a) $\xi /a_\mathrm{IB}=-1$ and b) $\xi /a_\mathrm{IB}=0$. All results are obtained using $V_{\rm G}$ and the colors indicate the same values as in FIG.~\ref{fig:Binding_energy}. As expected the separation of the impurities increases with the effective range of the potential and also for lighter impurities. The bipolaron binding energy decreases with increasing effective range and for lighter impurities. }}
	\label{fig:Energy_pol_size_eff} 
\end{figure}

{\color{black} In FIG.~\ref{fig:comparison} we show the bipolaron energy for the $M = \infty$ case for two different impurity-boson interaction potentials. Namely, we compare a Gaussian potential with the soft van-der-Waals potential. Here we either fix the effective range or the scattering length. We find that on either side of the scattering resonance and at the resonance the exact shape of the potential does not influence the results considerably and the exact choice of the underlying potential is not highly relevant for the obtained bipolaron energy. Interestingly, the bipolaron binding energy increases with a decreasing effective range. In FIG.~\ref{fig:comparison} d) we show the binding energy across the resonance for fixed effective range and we can see that the binding energy across the resonance is smooth.

In FIG.~\ref{fig:Energy_pol_size_eff} we show the separation of the impurities and the bipolaron binding energy for different mass ratios as a function of the effective range for a) $\xi/a_\mathrm{IB} =-1$ and b) $\xi/a_\mathrm{IB} =0$. As in the infinite mass case we observe, that energy decreases with the effective range. Additionally we see that impurity separation defined trough $\langle |\mathbf{\hat{R}}| \rangle = 4 \pi \int^\infty_0 \mathrm{d}{R}\, R |u(R)|^2$ stays of the order of the effective range of the underlying interaction potential.}

As mentioned earlier, it is essential to compare the localization of this bound state to the range of the {\color{black} direct} impurity-impurity interaction. 
Here we note that, on average, the separation is much greater than the range of the {\color{black} direct} impurity-impurity interaction, which in this case is actually the important length scale since after integrating out the bosonic degrees of freedom, we have reduced the problem effectively to a single particle scattering problem, with two competing length scales. To make this statement a bit more quantitative, we compare some characteristic effective ranges. 
Considering a microscopic van-der-Waals interaction the effective range is given by $r_0 \sim (C_6 m/m_e)^{1/4}a_0$ \cite{Pethick2008} , where $a_0$ is the Bohr radius and typical values are $C_6\sim10^3$ and $m/m_e\sim10^4-10^5$, which gives the estimate $r_0\sim 100a_0$. To put this into context, one can estimate the healing length in terms of the Bohr radius for typical experimental values (see for example \cite{Yan2019}), which leads to
 $\xi \sim 10^7-10^{10}a_0$. Hence the effective range of the {\color{black} direct} impurity-impurity interaction 
 typically a fraction of the range of the effective interaction potential mediated by the condensate. {\color{black} In FIG.~\ref{fig:Energy_pol_size_eff} it can also be clearly seen that the separation of the impurities is much larger than the effective range of the direct impurity impurity interaction. }
In the case of large impurity-bath interactions, the {\color{black} direct} impurity-impurity interaction can no longer be neglected, and few-body physics, like the bound state between the two impurities due to direct impurity-impurity interaction, can become 
 relevant. 
 %We also note that it might be interesting to explore the pair formation in greater detail and explore possible many-body effects between many impurities. The two-dimensional case should be of considerable  interest since here since an arbitrarily weak attractive potential the impurities is sufficient for the bound state formation.

\section{Conclusion}
We have presented an approach to the ground-state interaction of two impurities immersed into a three dimensional Bose gas capable of taking the boson-boson interaction into account. We started by showing that linearization efforts and the resulting Fr\"ohlich Hamiltonian are inadequate to 
 fully describe the polaron interaction in a Bose gas. We also discussed how naive extensions of the Fr\"ohlich model are inadequate. We then outlined how these issues can be addressed using a mean-field treatment paired with the Born-Oppenheimer approximation. The Born-Oppenheimer approximation is valid for heavy impurities. While strictly speaking, the mean-field approximation neglects quantum corrections in the form of modified phonons completely, it is important to note that the bipolaron properties are determined by short-scale physics. {\color{black} Therefore, we do not expect the modified phonons, which arise when including quantum corrections to play a significant role in this regime.} We first minimized the mean-field energy functional, from which we extracted the interaction potential. Here, we compared our results to the Yukawa potential and the results obtained from the linearized model  with a Gaussian potential. We then calculated the bipolaron energy using the effective potential by solving the resulting radial Schr\"odinger equation.  A detailed comparison of 
the results presented here with other methods and especially with the quasi-exact quantum Monte Carlo method would be a very interesting direction for future work.
The work highlights the fundamental problems like diverging mediated interactions, associated with approaches based on linearization when studying the interplay of two impurities and shows a simple way of dealing with these shortcomings. We hope that the {methods presented will} serve as fertile ground to explore the bipolaron in and out of equilibrium in greater detail. 

\section*{Acknowledgments}
We want to thank Michael Fleischhauer and Martin Will for valuable discussions.
JJ \ is grateful  for support from EPSRC under Grant
EP/R513052/1.
This research was supported in part by the National Science Foundation under Grant No. NSF PHY-1748958.

\subsection{Breakdown of the extended Fr\"ohlich model}
 In this Appendix we show how applying the standard variational method to the extended Fr\"ohlich model for the bipolaron problem will yield unphysical results. 
 This occurs because the emerging bound state is populated by an infinite number of excitations, which leads to a diverging energy. Within the BO approximation it is indeed possible to predict the position of this resonance fully analytically. Our starting point is the extended Fr\"ohlich model  $\hat{\mathcal{H}} = \hat{\mathcal{H}}_\mathrm{F} +\hat{\mathcal{H}}_\mathrm{2ph}  $ \cite{Frohlich1954,Grusdt2017b}, adapted to the two impurity case, in the $M\rightarrow\infty$ limit where
\begin{multline} \label{fröhlich}
\hat{\mathcal{H}} = \sum_\bk \Omega_{k} \ad_\mathbf{k} \oa_\mathbf{k}  +2\gib \sqrt{\frac{n_0}{L^d}}\sum_{\mathbf{k}\neq 0 } W_{k} \cos\left(\frac{\mathbf{k}\cdot\mathbf{{R}}}{2}\right) \\\big(\ad_\mathbf{k}  + \oa_\mathbf{-k}  \big) + \frac{\gib}{L^d}\sum_{\bk,\bkd \neq 0} \cos\left(\frac{(\bkd -\bk)\cdot\mathbf{R}}{2}\right)\\\Big[\big(W_kW_{k'}+W^{-1}_k W^{-1}_{k'}\big)  \ad_\bk\oa_\bkd 
+ \frac{1}{2} \big(W_kW_{k'}-  W^{-1}_k W^{-1}_{k'} \big) \\\Big(\ad_\bk \ad_{-\bkd}  + \oa_{-\bk}\oa_{\bkd}\Big) \Big].
\end{multline}
with $W_k = \left(\frac{(\xi k)^2}{2+(\xi k)^2}\right)^{1/4}$. In the BO approximation the Hamiltonian is quadratic and can therefore be solved by a coherent sate
 ansatz  $|\{\alpha_k\}\rangle$ (see \cite{Shchadilova2016} for a detailed discussion in the case of a mobile single impurity).
Applying the coherent state ansatz one obtains after some algebra, that the $\{\alpha_k\}$ can be chosen to be real, symmetric in $\bf k$ and are determined by the following self-consistent equation 
\begin{multline}
    \alpha_k = -2\gib\sqrt{n_0}\frac{W_k\cos(\bk \cdot \mathbf{R}/2)}{\Omega_k}\\  -2\gib \frac{W_k\cos(\bk \cdot \mathbf{R}/2)}{\Omega_k}  \frac{1}{L^d}\sum_{\mathbf{k'}} W_{k'} \cos(\mathbf{k'} \cdot \mathbf{R}/2)\alpha_{k'}
\end{multline}
which can be easily resummed as a geometric series. This leads to the following 
$\mathbf{R}$-dependent part of the ground state energy in the thermodynamic limit 
\begin{align} \label{eq:En}
    E(R) = \frac{n_0}{\frac{1}{2\gib} + \frac{1}{(2\pi)^d}\int d^dk\, \frac{W_k^2}{\Omega_k}\cos^2(\bk \cdot \mathbf{R}/2) }.
\end{align}
The integral $\int d^dk\, \frac{W_k^2}{\Omega_k}\cos^2(\bk \cdot \mathbf{R}/2)$ can be solved analytically in 3D using dimensional regularisation and yields
\begin{multline}
     \frac{1}{(2\pi)^3}\int d^3k\, \frac{W_k^2}{\Omega_k}\cos^2(\mathbf{k}\cdot\mathbf{R}/2) = \\
\frac{m}{(2\pi)^2} \left(\frac{-\sqrt{2}\pi}{\xi}+{\pi}\frac{\exp(-\sqrt{2}R/\xi)}{R} \right)
\end{multline}
It is now easy to see that the energy diverges when the denominator in (\ref{eq:En}) is zero,
which does not only depend on the coupling $\gib$ between the impurities but also the separation $R$. This can be traced back to the accumulation of an infinite number of phonons in the bound state. 
This is an effect that in reality is balanced by phonon-phonon interaction. A similar effect is known from the quantum mechanical setting see \cite{Albeverio2004} and Appendix B, where the bound state formation leads to an infinite energy in the thermodynamic limit. We note that other approaches that rely on trial wave function that do not re-sum the whole scattering series will not encounter this divergence.

\subsection{Two stationary impurities in ideal Bose gas}
In this Appendix, we discuss two stationary impurities in an ideal Bose gas. The appeal here is that one can solve this model analytically and study the emergence of the bound state in more detail. We consider $N$ bosons interacting with two static impurities located at $\pm R/2$ The Hamiltonian can now be expressed as the sum of single-particle Hamiltonians
\begin{align}\label{eq:Ham_ap}
    \hat{H} = \sum_n \Big(\frac{\hat{P}_n^2}{2m} + \frac{2\pi}{m}a_\mathrm{IB}\left[V(\hat{\bf Q}_n-{\bf R}/2)+V(\hat{\bf Q}_n+{\bf R}/2)\right]\Big),
\end{align}
where the interaction potentials are to be understood as boundary conditions on the wave function \cite{Albeverio2004,Chin2010a} which we will specify below.
First we note that for the eigenvalue equation associated with (\ref{eq:Ham_ap}) the wave function factorises $\Phi(r_1,r_2,...,r_N)=\phi(r_1)\phi(r_2)...\phi(r_N)$ and $E=N\mathcal{E}$. It is therefore sufficient to solve the following eigenvalue problem
\begin{equation} \label{eq:free_ham}
    \frac{-\nabla^2}{2m}\phi(r)=\mathcal{E}\phi(r),
\end{equation}
subject to the boundary condition (see \cite{Chin2010a} for details on the pseudo potential in the context of ultra cold gases) 
\begin{equation} \label{eq:boundary}
    \lim_{r_\pm \rightarrow 0} \left(r_\pm \phi(r)+ a_\mathrm{IB} \partial_{r_\pm}(r_\pm \phi(r))  \right) = 0,
\end{equation}
with $r_\pm = |{\bf r} \pm {\bf R/2}|$. This potential is always attractive and hosts a bound state in the single particle case only on the right side of the Feshbach resonance. 
The general solution to (\ref{eq:free_ham}) in spherical coordinates is given by $G(r) = \frac{e^{i k r}}{4\pi r}$. It can now be shown \cite{Albeverio2004}, that any solution satisfying (\ref{eq:free_ham}) and (\ref{eq:boundary}) with $\mathrm{Im}\,k = \kappa >0$ is of the form $\phi(r)= A G(r_+)+ B G(r_-)$. From (\ref{eq:boundary}) it follows then immediately, 
\begin{equation}
    \frac{1}{a_\mathrm{IB}}R -\kappa R = \pm e^{-\kappa R}.
\end{equation}
This equation has at least one solution if $-1 < \frac{1}{a_\mathrm{IB}}R$. Hence independent of $a_\mathrm{IB}$, there is always at least one bound state as long as the impurities are close enough together. Thus we see that having two impurities serves to enhance the possibility of having a bound state. Indeed, the above treatment suggests that there will always be a bound state if the impurities are sufficiently close together. However, it should be noted that using an approach that involves separate pseudopotentials is only valid when the impurities are sufficiently well separated. We note that this result does not depend on the choice of the pseudopotential 
and is also recovered if one chooses other regularisation schemes.
\subsection{Solving the radial Schr\"odinger equation}
In this Appendix, we outline the numerical approach taken to solve the radial Schr\"odinger equation. Usually, the ground state of radial Schr\"odinger equations is found employing the shooting method \cite{Killingbeck1987}. In recent years the field of scientific machine learning has made large improvements, and it has been shown that neural networks can be used to solve differential equations by leveraging their property of being universal function approximators \cite{Lagaris1998,Lagaris2000}. Another related use employs a neural network as a variational wave function to minimize an energy functional. This has been shown to yield good results for the ground state and also the first excited state of the stationary Schr\"odinger equation in \cite{Li2021}. Here, we are going to combine these two approaches and minimize the energy functional of the radial Schr\"odinger equation with an additional penalty term to enforce the boundary condition $u(0) = 0$.
In practice this can be written as a minimization problem with loss $L$
\begin{align}
    &u(x) = \mathrm{net}(x),\\ \nonumber
    &L[u] = (u, \hat{H}u)/(u,u) + \alpha (u,u),
\end{align}
where $(.,.)$, denotes the standard scalar product, $\hat{H}=\frac{-1}{2M}\frac{\mathrm{d}^2}{\mathrm{dR^2}} + V_\mathrm{BP}(R) $ is the radial part of the Hamiltonian and $\alpha$ is a hyper-parameter, that will be chosen such that $\alpha \gg E_g$, which ensures, that $u(0)=0$ . In practice this is implemented using PyTorch and we note that the derivatives arising in $(u, \hat{H}u)$ can be calculated exactly using PyTorch's automatic differentiation package. For the presented results we used a shallow network with only one hidden layer and a width of $1000$. 

\bibliographystyle{apsrev4-1} % Tell bibtex which bibliography style to use
\bibliography{library.bib} % 
\end{document}